\documentclass[%
 reprint,
 amsmath,amssymb,
 aps,
]{revtex4-1}

\usepackage{graphicx}
\usepackage{dcolumn}
\usepackage{bm}

\begin{document}


\maketitle


\noindent
\textbf{Comment on ``A Nonholonomic Model of the Paul Trap''}\\

A recent article by Borisov \emph{et al.}~\cite{borisov2018nonholonomic} studies the motion of a rigid ball in a rotating-saddle trap. The authors claim that they derive a new equation of motion from the Lagrangian formalism, which is different from the one we obtained from the Newtonian formalism in our recent work~\cite{fan2017confining}. We show here that these two equations of motion are the same. In addition, besides the reduced spin frequency $\sqrt{g}\omega_{0}/a$ and the moment of inertia coefficient $I/ma^{2}$, the stability condition given by the article is independent of the ball radius---this result is incorrect. The mistake is due to the fact that the center of mass $\bm{x}$ and the contact point $\bm{x}_{p}$ are not distinguished in the explicit expression of the local normal vector $\bm{\gamma}$.



We follow the original nomenclature of the article in this comment. The authors derive two equations of motion $\frac{d}{dt}\big( \frac{\partial L}{\partial \dot{\bm{x}}}\big)-\frac{\partial L}{\partial \bm{x}}=\bm{\lambda}$ and $\frac{d}{dt}\big( \frac{\partial L}{\partial \dot{\bm{\omega}}}\big)-\bm{\omega}\times\frac{\partial L}{\partial \bm{\omega}}=a\bm{\gamma}\times\bm{\lambda}$ for the center-of-mass coordinate $\bm{x}$ and angular velocity $\bm{\omega}$, with the system Lagrangian being $L=\frac{1}{2}m\dot{\bm{x}}^{2}+m\dot{\bm{x}}\cdot(\bm{\Omega}\times\bm{x})+\frac{1}{2}I(\bm{\omega}+\bm{\Omega})^{2}-V(\bm{x})$, and the sum of gravitational and centrifugal potentials being $V(\bm{x})=mgx_{3}-\frac{1}{2}m\Omega^{2}(x_{1}^{2}+x_{2}^{2})$. The gradient of the potential thus can be written in the form of $ \partial V/\partial \bm{x}=m\bm{\Omega}\times(\bm{\Omega}\times\bm{x})-m\bm{g}$.
By using this fact, noticing $\dot{\bm{\Omega}}=0$, and eliminating the multiplier $\bm{\lambda}$ with the aforementioned two equations of motion as is done in Eq.~1.6 of Ref.~\cite{borisov2018nonholonomic}, the following equation is obtained
\begin{equation}
\begin{aligned}
    &I\dot{\bm{\omega}}-I\bm{\omega}\times\bm{\Omega}\\
    &\ \ \ \ =a\bm{\gamma}\times\bigg[ m\ddot{\bm{x}}+2m\bm{\Omega}\times\dot{\bm{x}}+m\bm{\Omega}\times(\bm{\Omega}\times\bm{x})-m\bm{g} \bigg]
\end{aligned}
\label{eq:EOM1}
\end{equation}
where $a$ is the radius of the ball, and $\bm{\gamma}$ is the local normal vector of the saddle surface. 

Under the same notation, the equation of motion we derived in our recent work (Eq. 11 in Ref.~\cite{fan2017confining}) is
\begin{equation}
\begin{aligned}
    &I\dot{\bm{\omega}}+a\dot{\bm{\gamma}}\times m \dot{\bm{x}}+a\bm{\gamma}\times m\ddot{\bm{x}}\\
    &\ \ =a\bm{\gamma}\times\bigg[ m\bm{g}-m\bm{\Omega}\times(\bm{\Omega}\times\bm{x}) -2m\bm{\Omega}\times\dot{\bm{x}} \bigg]+I\bm{\omega}\times\bm{\Omega}
\end{aligned}
\label{eq:EOM2}
\end{equation}
By comparing Eq.~\ref{eq:EOM1} with Eq.~\ref{eq:EOM2}, two differences can be found: (1) the terms containing vector $\bm{\gamma}$ have opposite signs---this is due to the fact that the definitions of the local normal vector $\bm{\gamma}$ utilized in two articles are different in signs; (2) an extra term $(a\dot{\bm{\gamma}}\times m \dot{\bm{x}})$ appears in Eq.~\ref{eq:EOM2}---but this term is actually zero, which becomes obvious by expressing $\bm{\gamma}$ in terms of the positions of center of mass and the contact point as $\bm{\gamma}=(\bm{x}-\bm{x}_{p})/a$, and noticing that the velocity of the contact point $\dot{\bm{x}}_{p}$ is always parallel to the velocity of the center of mass $\dot{\bm{x}}$, namely $\dot{\bm{x}}_{p}\times\dot{\bm{x}}=0$, as we have already pointed out in the paragraph after Eq. 8 in Ref.~\cite{fan2017confining}. Therefore the equation of motion derived by the article's authors has no difference with the one we derived in Ref.~\cite{fan2017confining}. 

The authors also arranged the equation of motion with tensor notation in the article (Eq. 1.7 in Ref.~\cite{borisov2018nonholonomic}). But by noticing the facts that $(\bm{\gamma} \otimes \bm{\gamma})\dot{\bm{\omega}}= \bm{\gamma}(\bm{\gamma}\cdot \dot{\bm{\omega}})$ and $\bm{\gamma}\cdot\dot{\bm{\gamma}}=1/2\ \mathrm{d}(\bm{\gamma}^{2})/\mathrm{d}t=0$, it is straightforward to show that the tensor expression is still equivalent to the equation above. For the sake of clarity, we omit the detailed proof here~\footnote{there are mistakes that we believe to be typographical ones, e.g. a term in Eq. 1.9 in the article should be $d\bm{\gamma}\times[\bm{\Omega}\times (\bm{\Omega}\times\bm{x})]$.}.

Many previous works by the article's authors~\cite{borisov2002rolling} and other groups~\cite{fan2017confining, rueckner1995rotating} pointed out that the radius of the ball has an influence on the  dynamics of such systems. This can simply be understood by considering the motion of a frictionless rigid ball with radius $a$ on an uncompressed saddle surface $\Phi(\bm{x}_{p})=(x_{p1}^{2}-x_{p2}^{2})/x_{0}-x_{p3}=0$ rotating at frequency $\Omega$, where $\bm{x}_{p}$ is the contact point between the ball and the saddle surface. In this circumstance, the center of the rigid ball is elevated from the saddle surface $\bm{x}_{p}$ up to $\bm{x}=\bm{x}_{p}+\bm{\gamma}a$ due to the finite radius. As a consequence, the center of mass is now constrained on a new compressed saddle plane $\frac{x_{1}^{2}}{x_{0}-2a}-\frac{x_{2}^{2}}{x_{0}+2a}=x_{3}-a$ with aspect ratio $b=(x_{0}-2a)/(x_{0}+2a)<1$ up to the leading order of $\bm{\gamma}\cong \big(-2\frac{x_{p1}}{x_{0}},2\frac{x_{p2}}{x_{0}},1-2(\frac{x_{p1}}{x_{0}})^{2}-2(\frac{x_{p2}}{x_{0}})^{2} \big)$. Since the spinning of the ball is not coupled to its translation in the absence of friction, the system can be viewed as an effective mass point moving on the new saddle plane. As such, the stability condition of the saddle trap is given by $\Omega>\sqrt{2g/(x_{0}-2a)}$ according to previous researches on mass-point model (e.g. Ref.~\cite{kirillov2013exceptional,bottema1976stability}).

However, in the stability conditions given by the article's authors (Section 4 of Ref.~\cite{borisov2018nonholonomic}), the ball radius only enters the reduced spin frequency $\sqrt{g}\omega_{0}/a$ and the moment of inertia coefficient $I/ma^{2}$---the terms that are related to the spinning of the ball. This does not capture the fact that the elevation of center of mass due to the size of the ball modifies the stability, either when the ball is rolling or slipping on the saddle. The most obvious mistake the authors make is that the coordinates $\bm{x}_{p}$ and $\bm{x}$ are not distinguished in the vector field $\bm{\gamma}$. For instance, in Eq. 1.4 and 1.5 of Ref.~\cite{borisov2018nonholonomic}, the normal vector is the one corresponding to the center of mass plane, whereas the normal vector in the constraint Eq. 1.2 is the one corresponding to the contact point plane. These two local normal vectors are different. But as the radius $a\rightarrow 0$, we have $\bm{x}_{p}\rightarrow \bm{x}$. This is why in the limit of zero moment of inertia and spin angular velocity ($\omega_{0}=0, d=1$), the conclusion in the article (Case 1 in Section 4) happens to be consist with the mass-point model.

To conclude, although the correct equation of motion for a rigid ball in a rotating-saddle trap is obtained by the article's authors using Lagrange formalism, the attempt to study arbitrary aspect ratios of the saddle and arbitrary spin angular velocities of the ball does not appear valid. The correct extension of work~\cite{fan2017confining}, which studies unity aspect ratio of the saddle ($b=1$) and synchronized spinning of the ball ($\omega_{0}=0$), are warranted to further reveal the nonlinear nature of the system and clarify the phenomena such as high-speed instability.
\\\\
\noindent
Wenkai Fan\\
{\small Department of Physics, Duke University, Durham, North Carolina, USA, 27710\par}
\noindent\\
Li Du\\
{\small Department of Physics, Massachusetts Institute of Technology, Cambridge, Massachusetts, USA, 02139\par}
\noindent\\
Sihui Wang\\
{\small School of Physics, Nanjing University, Nanjing, China, 210093\par}
\noindent\\
Date: 23 June 2018 
\ \\
\ \\
\ \\
\ \\
\ \\
\ \\
\ \\
\ \\

\bibliography{main.bib}
\ \\
\ \\
\ \\
\ \\
\ \\
\ \\
\ \\
\ \\
\ \\
\ \\
\ \\
\ \\
\ \\
\ \\
\ \\
\ \\
\ \\
\ \\
\ \\
\ \\
\ \\
\ \\
\ \\
\ \\
\ \\
\ \\
\ \\
\ \\
\ \\
\ \\
\ \\
\ \\
\ \\
\ \\
\ \\
\ \\
\ \\
\ \\
\ \\
\ \\
\ \\
\ \\
\ \\
\ \\
\ \\
\ \\
\ \\
\ \\
\ \\
\ \\
\ \\
\ \\
\ \\
\ \\
\ \\
\ \\
\end{document}